\documentclass[aps,prb,twocolumn,groupedaddress,showpacs,notitlepage]{revtex4-1}
\usepackage{graphicx}
\usepackage{amsmath,amssymb}
\usepackage{color}
\usepackage{subfigure}
\newcommand{\hdag}{X^{\dagger}}
\newcommand{\hnod}{X^{\phantom{\dagger}}}

\newcommand{\adag}{a^{\dagger}}
\newcommand{\anod}{a^{\phantom{\dagger}}}

\begin{document}

\title{Excitonic Quasiparticles in a Spin-Orbit Mott Insulator } 

\author
{Jungho Kim,$^{1}$ M. Daghofer, $^{2}$ A. H. Said, $^{1}$ T. Gog,$^{1}$ J. van den Brink,$^{2}$
G. Khaliullin,$^{3}$ B. J. Kim$^{3,4,*}$}
\vspace{10 pt}
\address{$^1$Advanced Photon Source, Argonne National Laboratory, Argonne, Illinois 60439, USA}
\address{$^2$Institute for Theoretical Solid Sate Physics, IFW Dresden, Helmholtzstr. 20, 01069 Dresden, Germany}
\address{$^3$Max Planck Institute for Solid State Research, Heisenbergstra\ssße 1, D-70569 Stuttgart, Germany}
\address{$^4$Materials Science Division, Argonne National Laboratory, Argonne, IL 60439, USA}

\maketitle

{\bf
In condensed matter systems, out of a large number of interacting degrees of freedom emerge weakly coupled particles, in terms of which most physical properties are described. For example, Landau quasiparticles\cite{1} (QP) determine all electronic properties of a normal metal. The lack of identification of such QPs is a major barrier for understanding myriad exotic properties of correlated electrons, such as unconventional superconductivity\cite{2} and non-Fermi liquid behaviors.\cite{3} Here, we report the observation of a composite particle in a Mott insulator Sr$_2$IrO$_4$---an exciton dressed with magnons—--that propagates with the canonical characteristics of a QP: a finite QP residue and a lifetime longer than the hopping time scale. The dynamics of this charge-neutral bosonic excitation mirrors the fundamental process of the analogous one-hole propagation in the background of ordered spins,\cite{4} for which a well-defined QP has never been observed. The much narrower linewidth of the exciton reveals the same intrinsic dynamics that is obscured for the hole and is intimately related to the mechanism of high temperature superconductivity.    
}

The dynamics of a single hole doped into a Mott insulator is one of the unresolved fundamental issues in the physics of high temperature superconductivity in the cuprates.\cite{4,5,6} In one picture, the single hole forms a coherent quasiparticle (QP) propagating in a medium of ordered spins, which strongly renormalizes the dispersion relations and the mutual interactions of the holes. This spin-polaron picture, supported by numerical approaches, such as quantum Monte-Carlo,\cite{7} exact diagonalization\cite{8} and self-consistent Born approximation (SCBA),\cite{5} naturally connects to the pairing mechanism of pre-existing QPs glued by retarded bosons, analogous to Bardeen-Cooper-Schrieffer mechanism for conventional superconductivity. In a contrasting picture, the phenomenological absence of QPs in angle-resolved photoemission (ARPES)\cite{9,10,11} and scanning tunneling spectra\cite{12} is argued as due to orthogonality catastrophe,\cite{13} spin-charge separation,\cite{14} and/or localization effects,\cite{15} reflecting the unconventional nature of two-dimensional, strongly interacting fermions, which may support more exotic mechanisms of superconductivity such as the resonating valence bond theory proposed by Anderson.\cite{16} 

Motivated by the recent discovery of a new pseudospin-1/2 Heisenberg antiferromagnet on a square lattice, Sr$_2$IrO$_4$,\cite{17,18,19,20,21,22} we demonstrate in this letter a novel experimental approach using resonant inelastic x-ray scattering (RIXS), which is a rapidly evolving tool especially well suited for such 5d transition-metal oxides, to address this longstanding problem. Based on its remarkable similarity to superconducting cuprates in structural,\cite{22} electronic,\cite{17} and magnetic aspects,\cite{20,21} superconductivity has been predicted in Sr$_2$IrO$_4$.\cite{23} Although the electron correlation strength in Sr$_2$IrO$_4$ has been under much debate,\cite{24,25,26,27,28} questioning the validity of classifying this compound as a Mott insulator, Sr$_2$IrO$_4$ shares the same phenomenology that the hole spectral function as measured by ARPES lacks a legitimate QP.\cite{17} This similarity suggests a common origin for the absence of a well-defined QP in the ARPES spectra for two distinct but similar classes of materials. As we shall see, however, Sr$_2$IrO$_4$ supports a well-defined QP in another excitation channel, which RIXS is sensitive to and reflects the same dynamics as that of a hole, offering a novel route to studying elementary excitations in correlated oxides. 

\begin{figure}[t]
\hspace*{-0.1cm}\vspace*{-0.2cm}\centerline{\includegraphics[width=1\columnwidth,angle=0]{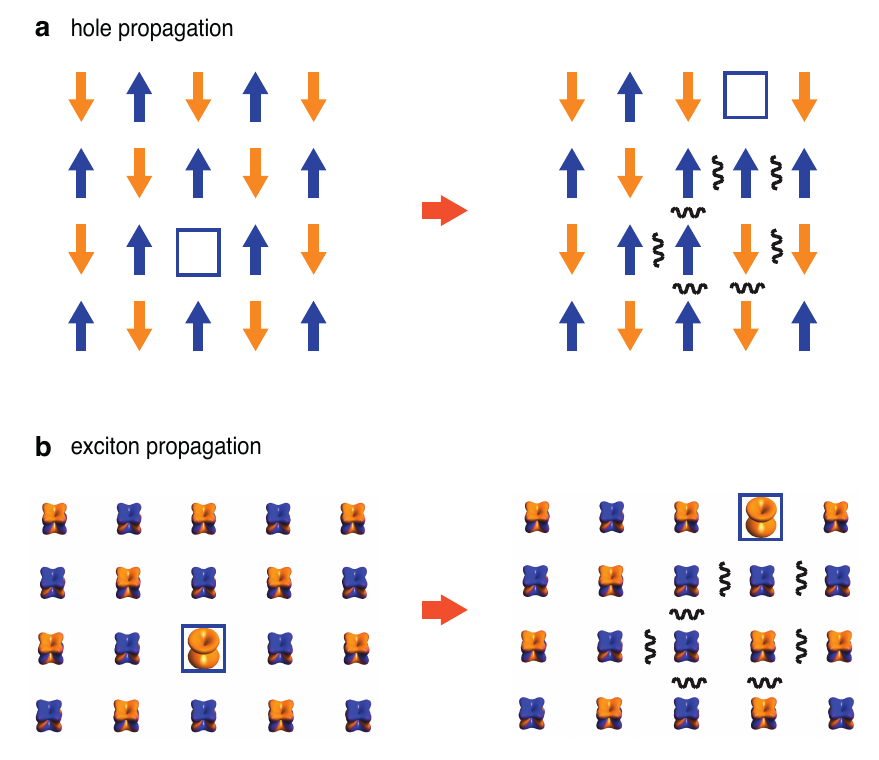}}
%\centerline{\includegraphics[width=4.5in,angle=-90]{fig1.eps}}%
\vspace*{-0.cm}%
%\centering\epsfig{file=fig1.eps,width=5cm,angle=-90}
%
\caption{{\bf Hole vs. exciton propagation in an antiferromagnetic background.} {\bf a}, a moving hole (blue square) creates a string of excited spins. Black wavy lines indicate pairs of `misaligned' spins. {\bf b}, an analogous exciton hopping.}\label{fig:fig1}
\end{figure}

In an earlier RIXS study of Sr$_2$IrO$_4$, dispersive d-d excitations (or excitons) across the spin-orbit coupling split levels (see Fig.~2a) have been identified and it has been shown that their dispersions can be understood in close analogy to the single hole problem.\cite{20} Figure 1 depicts the hole-vs.-exciton analogy: a ‘foreign’ object injected into a quantum Heisenberg antiferromagnet, be it a hole or an exciton, creates a string of flipped spins along its hopping path. This analogy, in principle, suggests that the dynamics of a particle moving in a magnetic medium can also be studied using the exciton. However, the energy resolution of RIXS used in the earlier study ($\approx$130 meV) was insufficient to resolve the dispersion and the intrinsic linewidth of the two exciton modes associated with the two pairs of Kramers doublets in the j=3/2 manifold (Fig.~2a). In this Letter, we exploit the different orbital symmetries of the two exciton modes to selectively probe each mode, and with the much improved energy resolution ($\approx$30 meV) offered by RIXS after recent developments, reveal their full dynamics.

\begin{figure}[t]
\hspace*{-0.1cm}\vspace*{-0.2cm}\centerline{\includegraphics[width=1\columnwidth,angle=0]{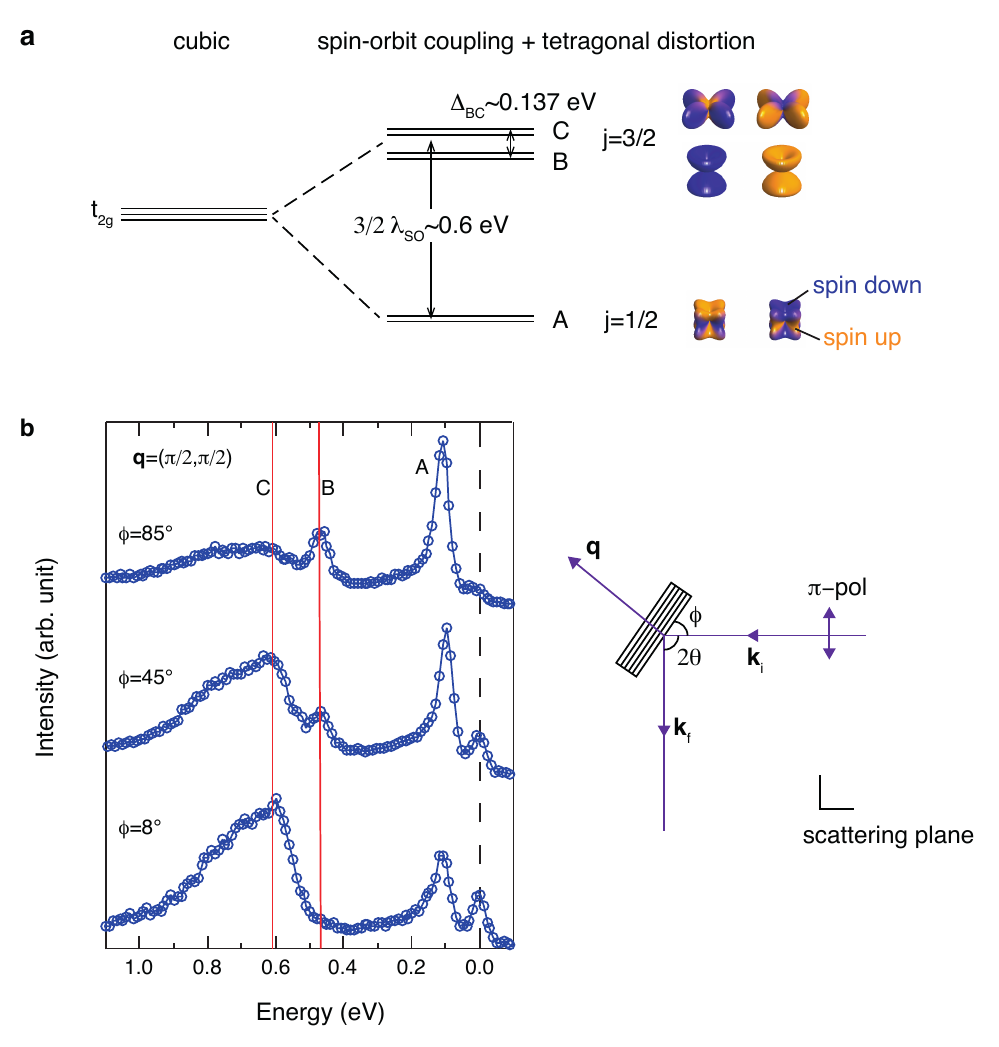}}
%\centerline{\includegraphics[width=4.5in,angle=-90]{fig1.eps}}%
\vspace*{-0.cm}%
%\centering\epsfig{file=fig1.eps,width=5cm,angle=-90}
%
\caption{{\bf RIXS spectra of Sr$_2$IrO$_4$.} {\bf a}, the spin-orbital level scheme of the three Kramers pairs and their orbital shapes in Sr$_2$IrO$_4$. The spin-orbit entangled nature of these quantum states is illustrated with colors; orange (blue) represent spin up (down) projection. {\bf b}, Spectrum at {\bf q}=($\pi$/2, $\pi$/2) measured at three different x-ray incident angle $\phi$. A, B, and C denote the energy positions of the three Kramers pairs illustrated in {\bf a}. The scattering angle 2$\theta$ was kept within 5 degrees from 90$^\circ$.}
\end{figure}

Figure 2b shows the RIXS spectra measured at three different incident angles $\phi$ of x-ray while fixing momentum transfer at {\bf q}=($\pi$/2, $\pi$/2). The two exciton modes, labeled as peaks B and C, show strong modulations in intensity as a function of φ through the change in the incident and outgoing x-ray polarizations relative to the sample surface and thereby the RIXS matrix elements. In particular, the peak B is strongly enhanced (completely suppressed) by tuning $\phi$ to normal (grazing) incidence geometry.  This strong matrix element effect enables selective mapping of B and C modes, as shown in Fig. 3a. While both B and C modes display rather similar dispersions, the B mode has much narrower linewidth (see Fig.~4c) partly due to the fact that it has lower excitation energy and thus has less phase space to decay into. With the two exciton modes disentangled, our high-resolution measurement yields for the B mode a bandwidth of ≈112 meV. This finding is consistent with the expectation that the bandwidth is on the scale of a few times the antiferromagnetic exchange coupling $J$ of about 60 meV.\cite{20} The global topology of the dispersions with minimum at {\bf q}=($\pi$/2, $\pi$/2) and maximum at the Γ point precisely matches that measured for a hole in cuprates by ARPES,\cite{9} which strongly supports the hole-vs.-exciton analogy.

\begin{figure}[t]
\hspace*{-0.1cm}\vspace*{-0.2cm}\centerline{\includegraphics[width=1\columnwidth,angle=0]{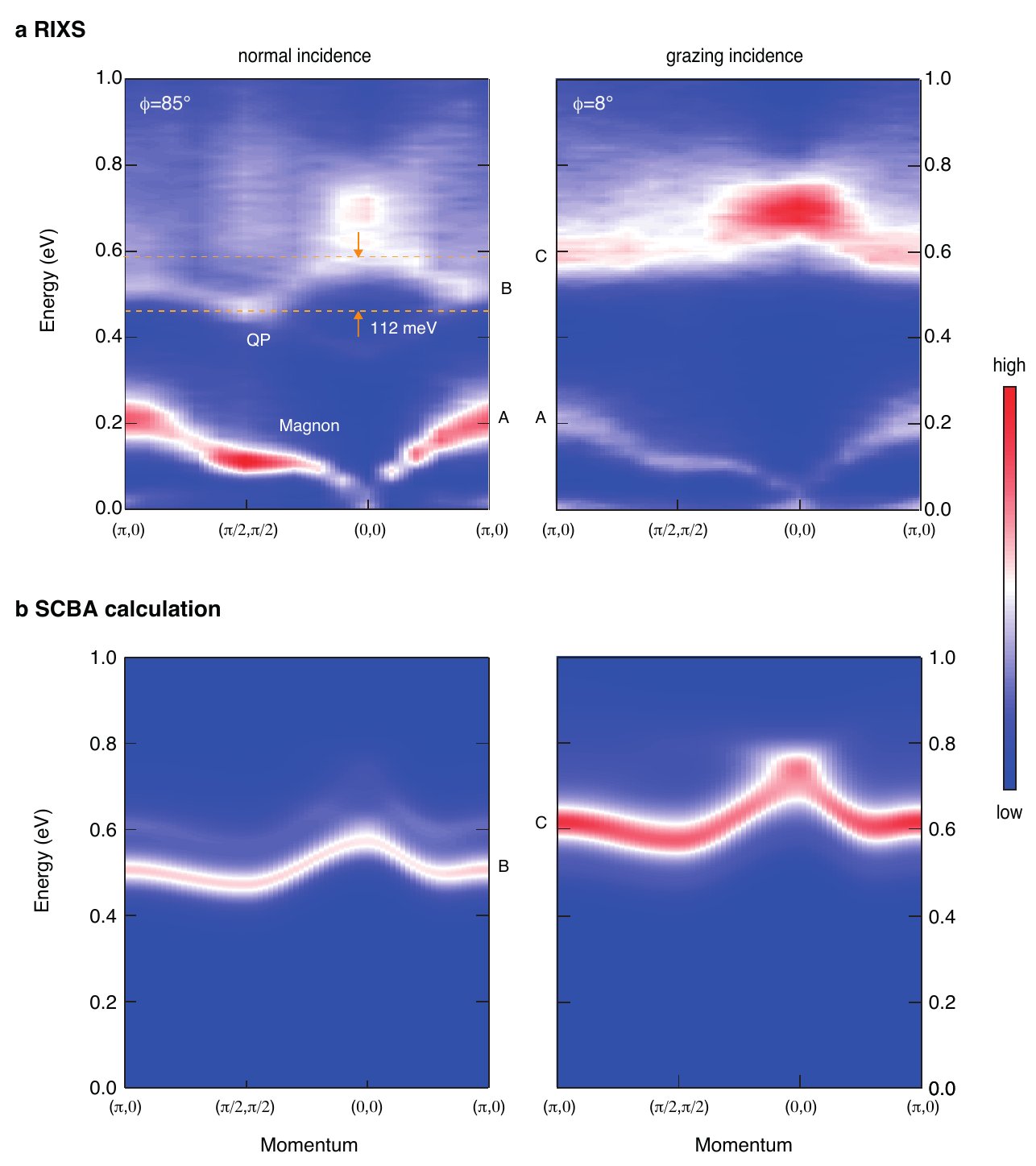}}
%\centerline{\includegraphics[width=4.5in,angle=-90]{fig1.eps}}%
\vspace*{-0.cm}%
%\centering\epsfig{file=fig1.eps,width=5cm,angle=-90}
%
\caption{{\bf Selective mapping of the two exciton modes and their comparison to SCBA calculations. } {\bf a}, Image plot of RIXS spectra measured along high symmetry lines in the normal and grazing incidence geometry. {\bf b}, SCBA calculations using the parameters $\lambda_{SO}$= 382 meV, $\Delta_{BC}$= 137 meV, $t_1$ = $J_1$/2 =30 meV, $t_2$ = $t_3$ = 7.6 meV, $J_1$ = 60 meV, $J_2$ = -20 meV, and $J_3$ = 15 meV. $t_{1,2,3}$ ($J_{1,2,3}$) denote first, second, and third nearest neighbor hoppings (magnetic couplings). $\lambda_{SO}$ and $\Delta_{BC}$ are defined in fig.~2. The spectral functions obtained by SCBA calculations are convoluted with a Lorentzian function with 5 meV width.}
\end{figure}

\begin{figure}[t]
\hspace*{-0.1cm}\vspace*{-0.2cm}\centerline{\includegraphics[width=1\columnwidth,angle=0]{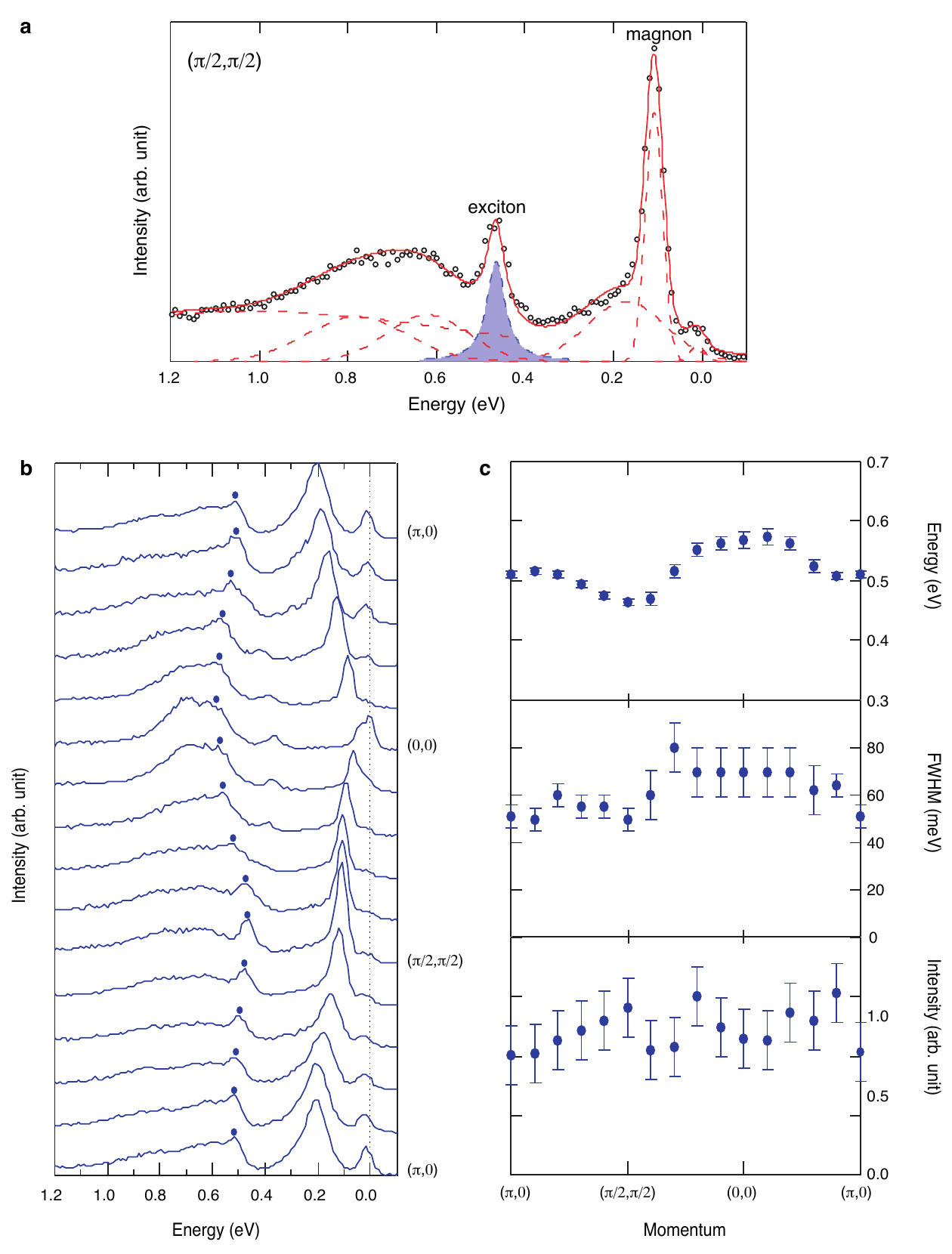}}
%\centerline{\includegraphics[width=4.5in,angle=-90]{fig1.eps}}%
\vspace*{-0.cm}%
%\centering\epsfig{file=fig1.eps,width=5cm,angle=-90}
%
\caption{{\bf Exciton dynamics in Sr$_2$IrO$_4$.} {\bf a}, RIXS spectrum at {\bf q}=($\pi$/2, $\pi$/2) (black open circles). The blue shaded peak corresponds to the exciton QP peak. The spectrum was fitted (red solid curve) using a Lorenzian lineshape for the exciton QP peak and Gaussian lineshapes for all other peaks (red dashed lines). The low energy features relative to exciton QP consist of elastic, single- and double-magnon peaks, and high-energy features the sum of C mode and background due to electron-hole continuum and incoherent part of B mode. {\bf b}, Stack plot of the image plot in Fig.~2a, left panel. Oval symbols mark the energy position of the QP. In addition to the QP, a small peak with the dispersion minimum at the $\Gamma$ point at E$\approx$0.37 eV is observed. A similar peak has been observed in a related material Na$_2$IrO$_3$ and attributed to a bound state at the edge of the particle-hole continuum.\cite{42}  {\bf c}, Energy, width, and intensity of the QP peaks along high symmetry lines extracted from peak fitting as exemplified in {\bf a}.
}
\end{figure}

\begin{figure}[t]
\hspace*{-0.1cm}\vspace*{-0.2cm}\centerline{\includegraphics[width=1\columnwidth,angle=0]{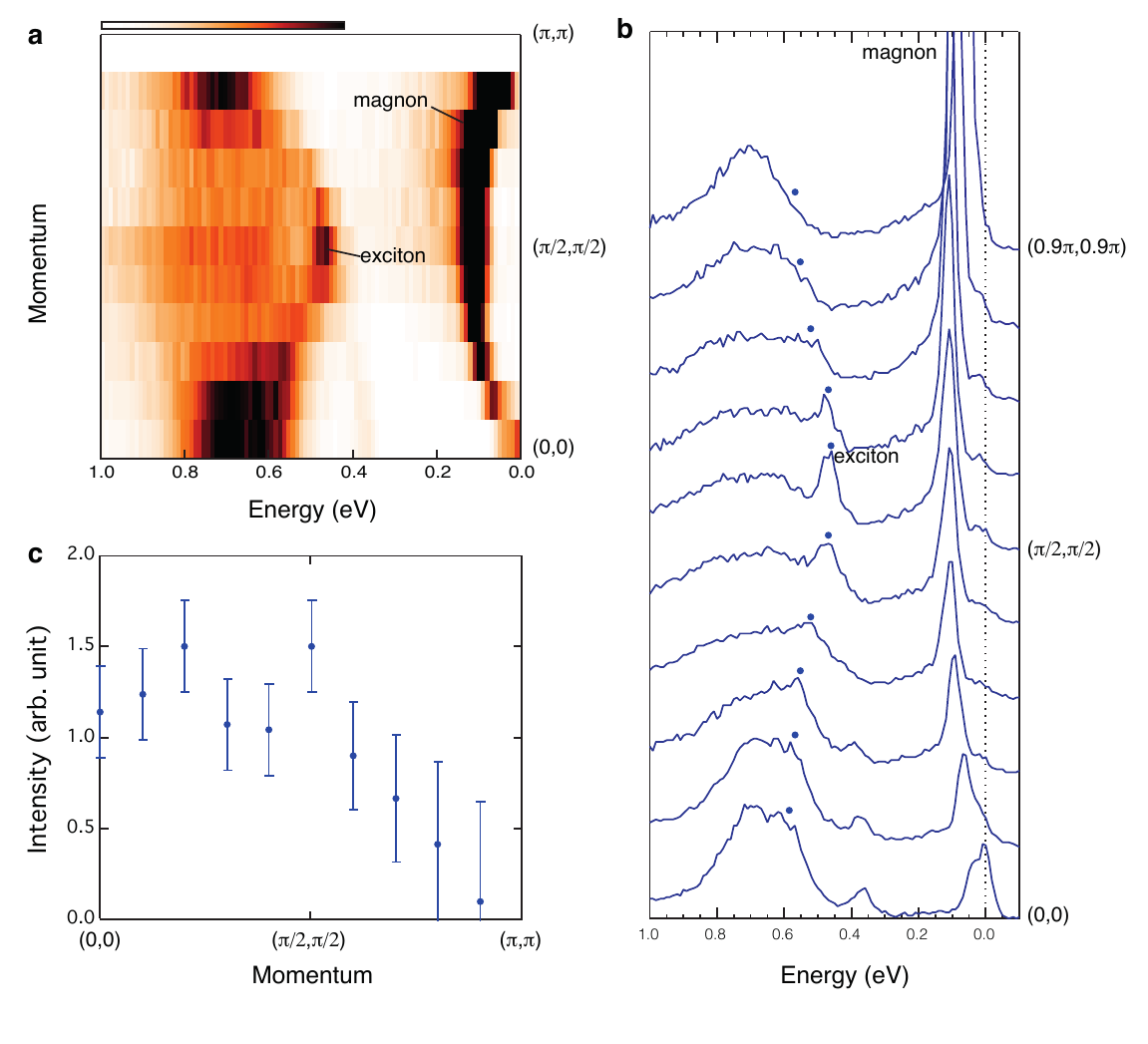}}
%\centerline{\includegraphics[width=4.5in,angle=-90]{fig1.eps}}%
\vspace*{-0.cm}%
%\centering\epsfig{file=fig1.eps,width=5cm,angle=-90}
%
\caption{{\bf Asymmetric exciton QP intensity. } {\bf a}, Image plot of RIXS spectra along (0,0)-($\pi$,$\pi$)  direction. {\bf b}, Stack plot of the image plot in {\bf a}. Oval symbols mark the energy position of the QP.  {\bf c}, Intensity of the QP peaks along (0,0)-($\pi$,$\pi$)  direction. }
\end{figure}

We compare the experimental data to the spectral function of the effective $t$-$J$ model calculated within the SCBA (Fig.~3b). For details, see Supplementary Information and Ref. 20. The calculation yields two modes, one of which is predominantly from the Kramers doublet with quantum numbers $|$J=3/2, J$_z$=$\pm$3/2$\rangle$, while the other has mostly $|$J=3/2, J$_z$=$\pm$1/2$\rangle$ character. Through an explicit calculation of the RIXS matrix elements, we find that their dependences on $\phi$ are such that the J$_z$=$\pm$1/2 states, with more in-plane components, are higher in energy (in the hole picture) than the J$_z$=$\pm$3/2 states. This level scheme is opposite to expectations based on perfectly cubic or c-axis elongated oxygen octahedra, but agrees with the results from non-resonant inelastic x-ray scattering,\cite{29} electron spin resonance\cite{30} and quantum chemistry calculations.\cite{31} Taking a crystal field splitting $\Delta_{BC}$ of 137 meV and fixing all other parameters to values inferred from independent studies, the effective $t$-$J$ model reproduces the gross features of the experimental spectra. Details are improved by including further neighboring hoppings, which are expected to be significant from sizable further neighbor magnetic couplings. With the parameters given in the Fig.~3b caption, we find an excellent agreement with the data in terms of the polarization dependence and the dispersion relations, confirming that the exciton dynamics is essentially captured by the effective $t$-$J$ model.

Having justified the hole-vs.-exciton analogy, we now bring to light the key observation from the exciton spectra. The energy distribution curves, measured in the normal incidence geometry to highlight the B mode, reveal a very sharp exciton peak, most prominent at {\bf q}=($\pi$/2, $\pi$/2) (Fig.~4a) and resolved throughout most part of the Brillouin zone (Fig.~4b). We use a phenomenological Lorentzian lineshape to fit the spectra to extract the peak energy, width, and intensity, which are summarized in Fig.~4c. Remarkably, the peak width is as narrow as $\approx$50 meV (of which 30 meV is contributed by the experimental resolution), much narrower than that of the sharpest peak ($\sim$200 meV) in the hole spectral function measured by ARPES for the same material.\cite{17,32} The peak width is also much smaller than its total bandwidth ($\approx$112 meV), which establishes the exciton as a propagating mode in a solid, or a QP. More importantly, our observation of an excitonic QP establishes hard evidence that a particle can propagate coherently through a quantum antiferromagnet. This raises a fundamental question: why is a QP absent for the single hole excitation?

Thus far we have focused on the similarity between the dynamics of a hole and an exciton. Let us now discuss some important differences. A hole is a charge monopole and its sudden creation in the ARPES process leads to deformation of the surrounding ionic oxygen cage, which results in a strong hole-lattice coupling detrimental for the hole propagation. Further, a hole interacts with charged impurities always present and poorly screened in an insulator. Both of these effects have been shown to strongly damp or wash out sharp QPs,\cite{33,34,35} thereby significantly redistributing the hole spectral function. In fact, most likely for these reasons, for a quasi-one-dimensional system Sr$_2$CuO$_3$ in which the phenomenon of spin-charge separation is established,\cite{36} ARPES measures much-broadened spectra\cite{37,38} of the theoretically predicted sharp edge-singularity in the exactly soluble model. By contrast, a charge-neutral exciton with a quadrupole moment should couple much more weakly to the lattice and is not subject to long-range Coulomb forces due to impurities. Thus, an exciton avoids these `side effects' and reveals the intrinsic dynamics of the $t$-$J$ model that have remained elusive for the past several decades. 

Our central message is that the unprecedentedly narrow linewidth of the QP allows a direct access into the nature of a QP living in the background of ordered spins, and thus that it allows direct verification of a class of theories that predict a finite quasiparticle residue in the Mott insulating phase, which has thus far remained inconclusive in the apparent absence of a QP in the hole channel. For instance, the observed exciton intensity can be compared with that calculated in theoretical models on a quantitative level. Figures 5a and 5b show the exciton dispersion along (0,0)-($\pi$,$\pi$) direction. The dispersion is symmetric with respect to ($\pi$/2,$\pi$/2) point on the magnetic zone boundary, but the intensities inside and outside of the magnetic zone differ significantly (Fig.~5c); the exciton QP displays a steep drop in intensity as ($\pi$/2,$\pi$/2) is crossed. This is a generic feature predicted for a hole in a $t$-$J$ model within a SCBA approach\cite{39} and indirectly inferred from ARPES measurements on cuprates,\cite{40} which possibly accounts for the Fermi arc observed in the doped case as due to the strongly suppressed shadow band intensity. Although some deviations from the cuprate physics might have been expected on general grounds due to fundamentally different nature of a hole and an exciton, and material specific differences between Sr$_2$IrO$_4$ and cuprates, the striking agreement down to a level of fine details with theories constructed for hole dynamics in cuprates demonstrates excellent parallel between the low-energy physics of Sr$_2$IrO$_4$ and the cuprates. 

\vspace{5 mm}
\noindent
{\bf Methods}
Single crystals of Sr$_2$IrO$_4$ were grown by the flux method. The sample was mounted in a displex closed-cycle cryostat and measured at 15 K. The RIXS measurements were performed using the MERIX spectrometer at the 30-ID beamline\cite{41} of the Advanced Photon Source. X-rays were monochromatized to a bandwidth of 15 meV, and focused to have a beam size of 45(H)$\times$30(V) $\mu$m$^2$. A horizontal scattering geometry was used with the π incident photon polarization. A Si (844) diced spherical analyzer with 4 inch radius and a position-sensitive silicon microstrip detector were used in the Rowland geometry. The overall energy resolution of the MERIX spectrometer at the Ir L$_3$ edge was 30 meV, as determined from the full-width-half-maximum of the elastic peak.

\newpage

\noindent {\large\bf Supplementary Note 1}\\ \\
{\bf The effective $t$-$J$ model and the calculation of exciton spectra within self-consistent Born approximation}\\ \\
The theoretical modeling is based on the observation that propagation
of an orbital excitation in a Mott insulator can be mapped onto a
single hole moving in the system,~\cite{mapiri,maptheory} as illustrated in
Fig.~1. We extend here the approach used in Ref.~\onlinecite{mapiri} by employing a more accurate technique, the self-consistent Born
approximation (SCBA).~\cite{scba} In contrast to the second-order
perturbation theory used previously,~\cite{mapiri} which is valid for 
exciton hopping small compared to magnetic interactions, it contains
diagrams to all orders of the hopping, i.e, in the
coupling between exciton motion and the magnetic background. Together
with the much increased experimental resolution, the use of the SCBA
rather than perturbation theory allows a more quantitative comparison
of theory and experiment. We find that a good description of the data
can be obtained with spin-orbit coupling $\lambda=382\;\textrm{meV}$, 
crystal-field splitting $\Delta=-188\;\textrm{meV}$ (in
the notation of Ref.~\onlinecite{george}), magnetic
couplings fitted to the magnon dispersion~\cite{mapiri} and by
allowing for longer-range hopping of the spin-orbit
excitation. As mentioned in the main text, these parameter values are
well supported by other experiments and ab initio methods 
-- with the exception of the longer-range hoppings, which can only be
investigated via precise measurement of the exciton propagation, as
done here.

The SCBA is a diagrammatic
approach that describes the antiferromagnetic (AFM) background with linear spin-wave
theory. The diagrams describing coupling of the moving hole or exciton to the (pseudo-)spin
background are those without crossing magnon lines, an approximation
that has been shown to be valid as the leading crossing diagrams drop
out for symmetry reasons.~\cite{nca} The method has been extensively used to
investigate hole motion in an AFM and has been used to discuss propagation of an orbital
excitation~\cite{scbaorb_2d}. In
contrast to second-order perturbation theory, it allows to describe
incoherent spectral weight as well as coherent quasi-particle motion. Nevertheless, it underestimates quantum
fluctuations of the magnetic background and higher-order vertex
corrections may become more relevant for a hopping of a similar size
as exchange coupling.

 \begin{figure}
 \setcounter{figure}{0}
 \renewcommand{\thefigure}{S\arabic{figure}}
 \subfigure[]{\includegraphics[width=0.49\columnwidth]{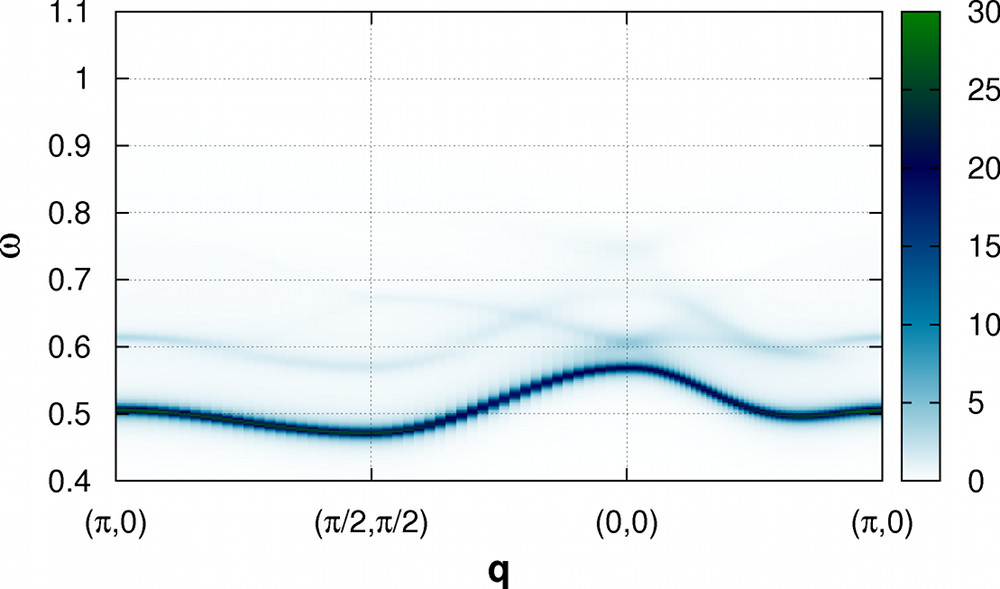}\label{fig:90}}
 \subfigure[]{\includegraphics[width=0.49\columnwidth]{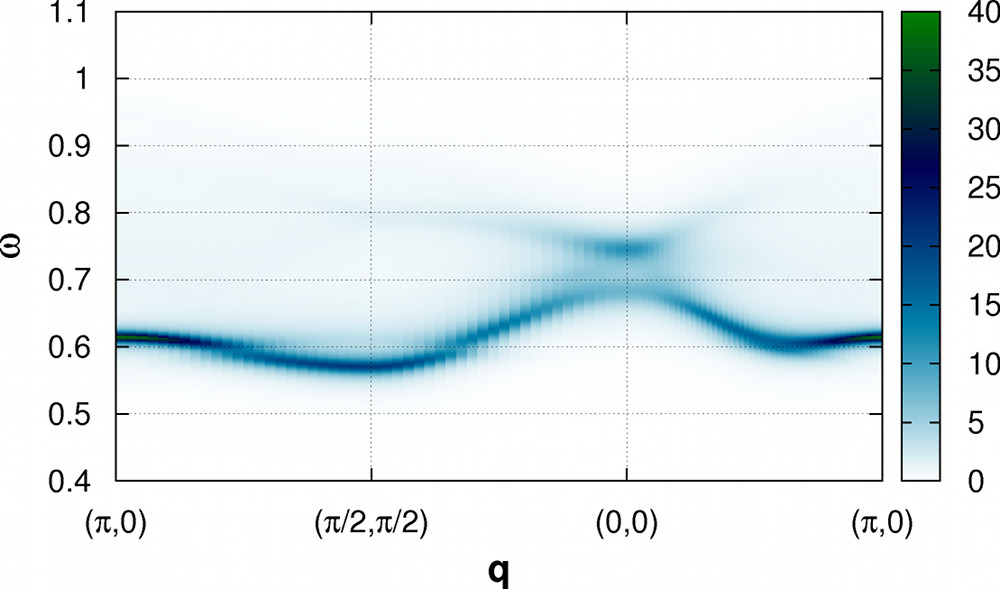}\label{fig:0}}\\
 \subfigure[]{\includegraphics[width=0.49\columnwidth]{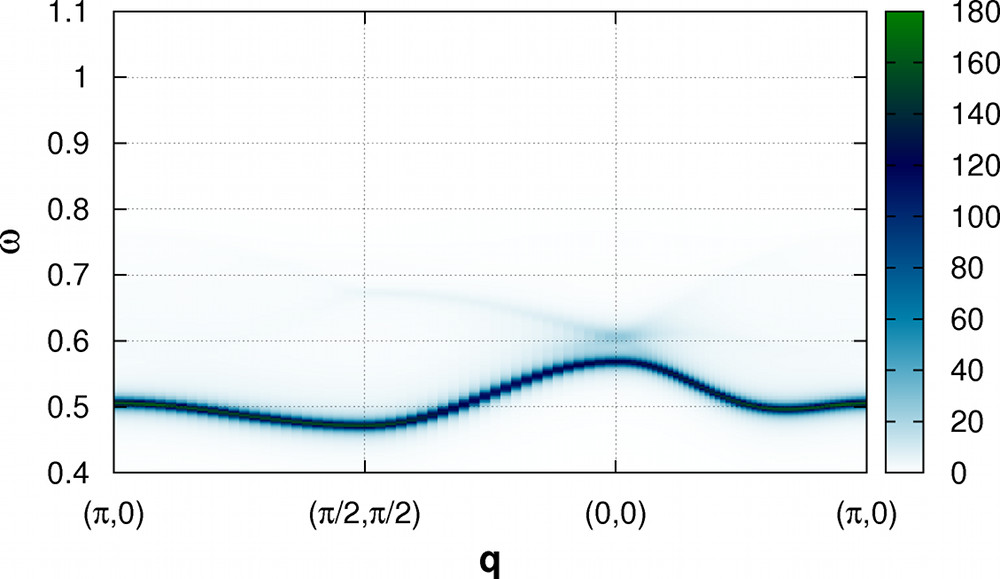}\label{fig:orbB}}
 \subfigure[]{\includegraphics[width=0.49\columnwidth]{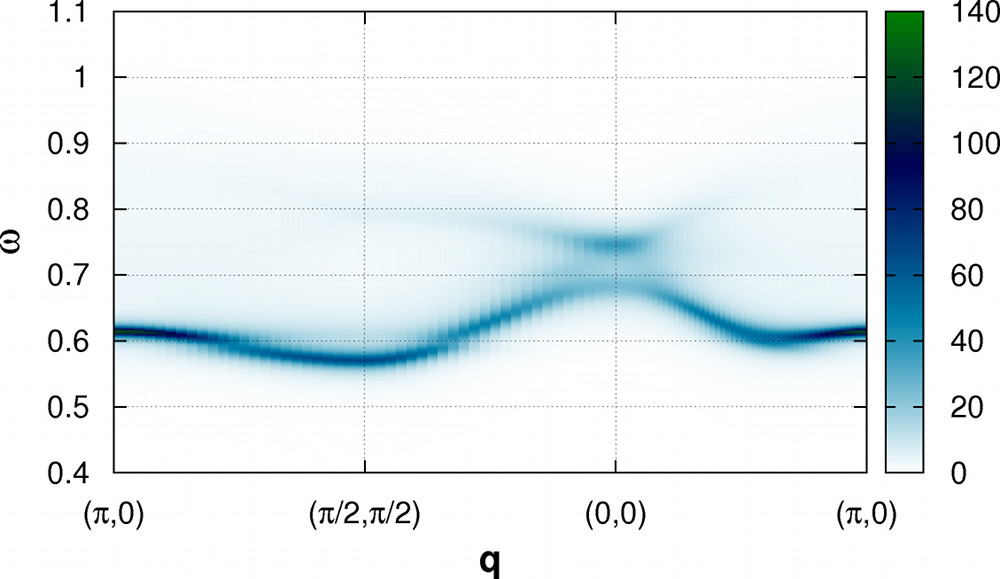}\label{fig:orbC}}\\
 \caption{Theoretical spectra for a scattering geoemtry with (a)
   $\phi$=$90^\circ$ and (b) $\phi$=$0^\circ$. In (c) and (d), the weight for the two
 doublets {\rm B} and C is shown, i.e. $-\mathcal{I} G^{{\rm BB}}$ in (c) and
 $-\mathcal{I} G^{{\rm {\rm CC}}}$ in (d). The parameters used were
 $\lambda=382\;\textrm{meV}$, $\Delta=-188\;\textrm{meV}$, $W_1=J_1/2=30 \;\textrm{meV}$,
 $W_2=W_3=7.6\;\textrm{meV}$, $J_1 = 60 \;\textrm{meV}$, $J_2 =
 -20 \;\textrm{meV}$, and $J_3 = 15 \;\textrm{meV}$. Imaginary part
 used in the SCBA was $\delta = 5\;\textrm{meV}$.}
 \end{figure}

The {\rm A}, {\rm B} and C modes all mix spin and
orbital of the $5d$ wave functions, depending on spin-orbit coupling
and crystal-field splitting $\Delta$ between the $xy$ and $xz/yz$
orbitals.  The low-energy {\rm A} states are then
given by $|{\rm A}_{\sigma}\rangle = \sin \theta |\tau=0;\sigma\rangle - \cos
\theta |\tau=\sigma;-\sigma\rangle$, where $\sigma=\pm 1$ denotes the
pseudospin and $\tau=0, \pm 1$ the orbital angular momentum and $\theta$
is given by $\tan 2 \theta  = 2\sqrt{2}\lambda/(\lambda-2\Delta)$.\cite{george} 
The higher-energy modes {\rm B} and {\rm C}, which we  focus on here, are 
given by 
\begin{align}\label{eq:BC_theta}
   |{\rm B}_\sigma\rangle &= |\tau=\sigma; \sigma\rangle, \quad\textrm{and}\\
   |{\rm C}_\sigma\rangle &= \cos \theta\ |\tau=0;\sigma\rangle + \sin \theta\ |\tau=\sigma;-\sigma\rangle\;.
\end{align}
The energy cost of exciting a hole from the ${\rm A}$ into the
${\rm B}$ or ${\rm C}$ doublets is given by 
\begin{align}\label{eq:E_BC}
  E^{{\rm C}} &=  \frac{3\lambda}{2}\sqrt{1- \frac{4\Delta}{9\lambda} + \frac{4\Delta^2}{9\lambda^2} }\nonumber\\
  E^{{\rm B}} &= \frac{\Delta}{2} + \frac{3\lambda}{4} + \frac{3\lambda}{4}\sqrt{1- \frac{4\Delta}{9\lambda} + \frac{4\Delta^2}{9\lambda^2} }\;.
\end{align}
The splitting between {\rm A}, {\rm B} and C modes observed in the RIXS data can
be reproduced by taking spin-orbit coupling
$\lambda=382\;\textrm{meV}$ and a  crystal-field splitting
$\Delta=-188\;\textrm{meV}$, yielding $\theta \approx 0.15\pi$. The sign of $\Delta$ is here
  chosen as in Ref.~\onlinecite{george}, i.e., it is positive for
  elongated octahedra and negative for shortened ones, if one considers a simple point-charge model. However, we find that in order to reproduce the order of the {\rm B} and
C modes as seen  in experiment, 
  and consistent with first-principle calculations~\cite{qu_chem} and other
  experiments~\cite{xray_esr}, $\Delta< 0$ has to be chosen  for Sr$_2$IrO$_4$. In the notation used in the main text, the
  resulting splitting $E^{\rm C}-E^{\rm B} = 
  \Delta_{\textrm{BC}} = 137.4\;\textrm{meV}$ between {\rm B}
and C levels has the sign opposite to that of $\Delta$, i.e., it is
here positive.

As discussed in Ref.~\onlinecite{mapiri}, the orbital
excitation moves via a superexchange process with the matrix element
\begin{align}\label{eq:eff_hopp}
W^{\beta\gamma}_{x/y} &= -\frac{2t_{{\rm AA}}}{U'}\left(\begin{array}{cc}
t_{BB}& \pm t_{{\rm BC}}\\
\pm t_{{\rm BC}}& t_{{\rm CC}}
\end{array} \right) = \\
= &-\frac{J}{2}\frac{U}{U'}\left(\begin{array}{cc}
t_{BB}/t_{{\rm AA}}& \pm t_{{\rm BC}}/t_{{\rm AA}}\\
\pm t_{{\rm BC}}/t_{{\rm AA}}& t_{{\rm CC}}/t_{{\rm AA}}
\end{array} \right) = -W_1\tau^{\alpha\beta}_{x/y}
\;\nonumber
\end{align}
along the $x$ and $y$ directions, respectively.
$t_{\alpha,\beta}$ are here the direct hopping
elements for the three Kramers-doublets {\rm A}, {\rm B}, and C. $U'$ is the
Coulomb repulsion between two electrons on the same site, but in
different orbitals. Since $SU(2)$ symmetry pseudospin excitations of
the {\rm A} sector indicate small Hund's rule coupling, it is not expected
to differ much from the intraorbital repulsion $U$ and we take here
$U'=U$.  The overall factor scaling the propagation is then
$W_1=\frac{J}{2}\approx 30\;\textrm{meV}$. To
  distinguish them more clearly from the bare hopping elements $t$ referring
  to electron motion, 
  effective exciton hoppings are denoted by $W$ in this supplementary
  material, following earlier work.~\cite{mapiri}
The bare hopping elements entering (\ref{eq:eff_hopp}) depend on the composition of the
states as
% \begin{align}
% t_{{\rm AA}} & = \frac{3}{4} - \frac{\cos 2\theta}{4} \nonumber \\
% t_{BB} & = \frac{1}{2} \label{eq:hopp}\\
% t_{{\rm CC}} & = \frac{3}{4} + \frac{\cos 2\theta}{4} \nonumber\\
% t_{{\rm BC}} & = \frac{\sin \theta}{2}\;. \nonumber
% \end{align}
\begin{align}
t_{{\rm AA}} & = \frac{3}{4} - \frac{\cos 2\theta}{4},\qquad
t_{{\rm BB}}  = \frac{1}{2} \label{eq:hopp}\\
t_{{\rm CC}} & = \frac{3}{4} + \frac{\cos 2\theta}{4},\qquad
t_{{\rm BC}}  = \frac{\sin \theta}{2}\;. \nonumber
\end{align}

As the excitonic hopping is a superexchange process, it scales with
the magnetic exchange $J$, see (\ref{eq:eff_hopp}). In
Sr$_{2}$IrO$_{4}$, nearest-neighbor exchange is here taken as
$J=60\;\textrm{meV}$. Longer-range couplings are also present and
sizable with $\approx 15 - 20\;\textrm{meV}$, analogously, we
introduce longer-range exciton hoppings $W^{\beta\gamma}_{2}$ and
$W^{\beta\gamma}_{3}$ for second and third neighbors. For simplicity,
they are assumed to be identical for the {\rm B} and C doublet with
$W^{{\rm BC}}_{2/3}=0$, leaving us with two parameters
$W^{{\rm BB}}_2=W^{{\rm CC}}_2=W_2$ and $W^{{\rm BB}}_3=W^{{\rm CC}}_3=W_3$. They 
are adjusted to fit the quasi-particle dispersion, like longer-range
hoppings in the case of  ARPES. Exciton hopping and magnetic
background together are thus described by an effective $t$-$J$ model,
where both hopping and magnetic exchange are included up to third
neighbors. In order to make the analogy to cuprates more transparent,
  $W_1$, $W_2$ and $W_3$ are denoted by $t_1$, $t_2$, and $t_3$ in the
  main text.
One difference to the $t$-$J$ model used to
describe hole motion in an antiferromagnet is the orbital index of the
exciton.  Moreover, hopping
$W$ and magnetic coupling $J$ are of the same order of magnitude, as
they share the same origin, superexchange. 

% Second- and third-neighbor hoppings $W_2$ and
% $W_3$ allow the
% exciton to move within one sublattice of the AFM background, and
% consequently do not disturb the the magnetic order. Nearest-neighbor
% hopping, in contrast, is connected to magnon creation/annihilation and
% can be written as 
% \begin{equation}
% H_{\textrm{kin}}=\sum_{\stackrel{\langle i,j\rangle \parallel x,y}{\alpha,\beta}} W^{\alpha\beta}_{x/y} \; X^{\dagger}_{i\alpha} 
% X^{\phantom{\dagger}}_{j\beta} \; (b^{\dag}_j + b^{\phantom{\dagger}}_i), 
% \label{eq:H}
% \end{equation}
% where  $X^{\phantom{\dagger}}_{i,\alpha}$ ($X^{\dag}_{i\alpha} $)
% annihilates (creates) a spin-orbit exciton on lattice site $i$ and
% with pseudoorbital index $\alpha = {\rm B},C$.  $b^{\dag}_i$ and $b_i$ create
% and annihilate magnons. $\langle i,j \rangle$ indicate
% nearest-neighbor bonds, directional and orbital dependence of
% $W^{\alpha\beta}_{x/y}$ are given in Eqs.~(\ref{eq:eff_hopp}) and
% (\ref{eq:hopp}).  Apart from the orbital index of the exciton, the problem of a
% particle moving an and AFM background has been
% extensively studied in the context of hole propagation in the $t$-$J$
% model and we accordingly proceed in an analogous manner.

In order to treat this problem, we follow a route extensively used to
describe hole propagation in the $t$-$J$ model, where the AFM
order is treated in linear spin-wave approximation. For the case 
with only nearest-neighbor terms and for a single excited-orbital
mode, it has been used to discuss propagation of an orbital
excitation~\cite{scbaorb_2d}. 
The
Hamiltonian is Fourier-transformed into momentum space and a
Bogoliubov transformation is applied to the magnetic part. The
magnetic energy becomes
\begin{align}\label{eq:Jq}
H_{\textrm{mag}}= \sum_{\bf q} \omega_{\bf q}^{\phantom{\dagger}}
\adag_{{\bf q}}\anod_{{\bf q}}
\end{align}
where $\anod_{{\bf q}}$ ($\adag_{{\bf q}}$) annihilates(creates)  a
magnon with momentum ${\bf q}$, with a  magnon energy
\begin{align}\label{eq:bog_omega}
\omega_{\bf q} &= \sqrt{{\rm A}_{\bf q}^2-{\rm B}_{\bf q}^2}\;,
\end{align}
% Like the Bogoliubov factors
% \begin{align}\label{eq:bog_uv}
% u_{\bf q} = \frac{1}{\sqrt{2}} \sqrt{\frac{{\rm A}_{\bf q}}{\omega_{\bf q}}+1}, \quad 
% v_{\bf q} = -\frac{\textrm{sgn} {\rm B}_{\bf q}}{\sqrt{2}}
% \sqrt{\frac{{\rm A}_{\bf q}}{\omega_{\bf q}}-1} \;,
% \end{align}
% $\omega_{\bf q}$ is 
and coefficients 
\begin{align}\label{eq:bog_AB}
{\rm A}_{\bf q} &= 2(J_1-J_2-J_3+J_2\cos q_x\cos q_y)+J_3(\cos 2q_x+\cos 2q_y), 
\nonumber\\ 
{\rm B}_{\bf q} &= J_1(\cos q_x+\cos q_y)\;.
\end{align}
The magnetic couplings are chosen as $J_1 = 60 \;\textrm{meV}$, $J_2 =
-20 \;\textrm{meV}$, and $J_3 = 15 \;\textrm{meV}$, in order to
describe the `magnon' excitations of the pseudospin-$1/2$ checkerboard
order.~\cite{mapiri} 
Nearest-neighbor hopping moves the exciton, i.e., the hole in the AFM
background from one sublattice of the AFM order to the other and
thereby couples to the magnetic background. It can be written as 
\begin{align}\label{eq:ham_kin}
H_{\textrm{NN}} =-\frac{zW_1}{\sqrt{N}} \sum_{{\bf k},{\bf q},\alpha,\beta} 
\Bigl[M^{\alpha\beta}_{{\bf k},{\bf q}}
\hdag_{\alpha{\bf k}}\hnod_{\beta{\bf k}-{\bf q}} \anod_{{\bf q}} + h.c. \Bigr]\;,
\end{align}
where $N$ is the number of sites and $z=4$ the coordination number.
$\hdag_{\alpha{\bf k}}$ and $\hnod_{\beta{\bf k}}$ create and
annihilate  the exciton at momentum ${\bf k}$ and orbital index
$\alpha$.
The orbital and direction dependence of Eq.~(\ref{eq:eff_hopp}) was
absorbed into the orbital-dependent vertex  
\begin{align}\label{eq:vertex}
M^{\alpha\beta}_{{\bf k},{\bf q}}=
|\tau^{\alpha\beta}|(u^{\phantom{\beta}}_{\bf q}\gamma^{\alpha\beta}_{{\bf k}-{\bf q}}
+ v^{\phantom{\beta}}_{\bf q}\gamma^{\alpha\beta}_{{\bf k}})
\end{align}
via 
\begin{align}
\gamma^{\alpha\beta}_{\bf k}=\frac{1}{2}\left(\begin{array}{cc}
(\cos k_x + \cos k_y)& (\cos k_x - \cos k_y)\\
(\cos k_x - \cos k_y)& (\cos k_x + \cos k_y)
\end{array} \right)\;;
\end{align}
Bogoliubov factors $u_{\bf q}$ and $v_{\bf q}$ are given by the
relations  
\begin{align}\label{eq:bog_uv}
u_{\bf q} = \frac{1}{\sqrt{2}} \sqrt{\frac{{\rm A}_{\bf q}}{\omega_{\bf q}}+1}, \quad 
v_{\bf q} = -\frac{\textrm{sgn} {\rm B}_{\bf q}}{\sqrt{2}}
\sqrt{\frac{{\rm A}_{\bf q}}{\omega_{\bf q}}-1}\;.
\end{align}

Second- and third-neighbor hoppings $W_2$ and
$W_3$, in contrast, allow the exciton to move within one sublattice of the AFM
background and consequently do not disturb the the magnetic
order. Together with the onsite energies $E^{\rm B}$ and $E^{\rm C}$ reflecting the energy cost of exciting a hole from the {\rm A} into the
{\rm B} or C doublets, see (\ref{eq:E_BC}), they add a `free' non-interacting
exciton Hamiltonian: ~\cite{scba_tp}
\begin{align}\label{eq:free}
H_{0} &= -\frac{1}{\sqrt{N}} \sum_{{\bf q},\alpha} \epsilon^\alpha_{\bf
  q}\hdag_{\alpha{\bf q}}\hnod_{\alpha{\bf q}}\quad \textrm{with} \\
%\epsilon^\alpha_{\bf  q} &= E^{\alpha} - z (W_2 \cos {\bf q}_x \cos {\bf q}_y + W_3 \gamma^{\alpha\alpha}_{2\bf q})\;.\nonumber
\epsilon^\alpha_{\bf  q} &= E^{\alpha} - 4W_2 \cos {\bf q}_x \cos
{\bf q}_y - 2W_3 ( \cos 2{\bf q}_x  + \cos 2{\bf q}_y).\nonumber
\end{align}

The full Hamiltonian $H_0+H_{\textrm{NN}}+H_{\textrm{mag}}$ is then
treated using the self-consistent Born approximation,~\cite{scba} a diagrammatic
approach where the self energy is expressed as
\begin{align}\label{eq:sigma}
\Sigma^{\alpha\beta}({\bf k},\omega)=-\frac{z^2W^2}{N} \sum_{{\bf q}, \gamma,\gamma'} 
M^{\alpha\gamma}_{{\bf k},{\bf q}}
G^{\gamma\gamma'}({\bf k}-{\bf q},\omega-\omega_{\bf q})
M^{\gamma'\beta}_{{\bf k},{\bf q}},
\end{align}
with the interacting Green's function conveniently expressed via its
inverse matrix $G^{-1}$ as 
\begin{align}\label{eq:green}
\left\{G^{-1}({\bf q},\omega)\right\}^{\alpha\alpha} &= \omega +i\delta- \epsilon^\alpha_{\bf  q} - \Sigma^{\alpha\alpha}({\bf q},\omega),\\
\left\{G^{-1}({\bf q},\omega)\right\}^{\alpha\beta\neq\alpha} &= -\Sigma^{\alpha\beta}({\bf q},\omega).\nonumber
\end{align}
The small imaginary part $i\delta$ was here chosen as
$\delta=5\;\textrm{meV}$. Many of the diagrams neglected in thin
non-crossing approximation drop out for the square-lattice N\'eel
state, however, it underestimates quantum fluctuations of the
background.

The resonant inelastic x-ray scattering (RIXS) spectra can be simulated from the spectral density $-\mathcal{I}
G^{\alpha\beta}({\bf q},\omega)$ with proper consideration of the factors arising from the scattering geometry. The RIXS intensity can be expressed as
\begin{align}\label{eq:weight}
I &= -|\beta|^2\mathcal{I} G^{{\rm BB}}({\bf q},\omega)-|\gamma|^2\mathcal{I}
G^{{\rm CC}}({\bf q},\omega)\\ \nonumber
&\quad -\beta^*\gamma \mathcal{I} G^{{\rm BC}}({\bf q},\omega)-\beta\gamma^* \mathcal{I} G^{{\rm CB}}({\bf q},\omega)
\end{align}
where coefficients $\beta$ and $\gamma$ are obtained in a similar manner as in
Ref.~\onlinecite{mapiri}, extended here to account for the deviations from the
cubic limit. They depend on incoming and outgoing polarizations
$\epsilon$ and $\epsilon'$ as well
as on the spin of the excited particle. As the spin does not enter the
calculation of the spectral weight and the outgoing polarization $\epsilon'$ is
not analyzed, we average over both. The matrix elements are given by
\begin{align}\label{eq:coeffs_bc}
\beta_{\sigma,\sigma} &= \frac{1}{2}\cos \theta_0 \cos (\theta-\theta_0)(Q_2+i\sigma T_z),\\
\beta_{\sigma,-\sigma} &= \frac{1}{2\sqrt{2}}\sin \theta_0 \cos(\theta-\theta_0)
\left[-\sigma (T_y+P_y)+i(T_x-P_x)\right],\\
\gamma_{\sigma,\sigma} &= \frac{\sin(2\theta_0-2\theta)}{4\sqrt{3}} (\sqrt{2}Q_1+Q_3)
+\frac{\sqrt{3}\sin 2\theta }{4}Q_3\nonumber\\
&\quad+ i\sigma\frac{1}{4}[\sin 2\theta - \sin(2\theta_0-2\theta)]P_z\\
\gamma_{\sigma,-\sigma} &= \frac{1}{2\sqrt{2}}\left[\sigma(T_y -\cos 2\theta
P_y) +i (T_x+\cos 2\theta P_x)\right]
\end{align}
where $\theta_0$ is the `mixing angle' in the cubic limit, which
applies to the oxygen levels that are not affected by octahedral
distortions and is given by $\tan 2 \theta_0  = 2\sqrt{2}$. Further, 
\begin{align} 
\vec{P} & = \epsilon'\times\epsilon\\
 Q_1 &= \sqrt{2/3}\epsilon\cdot\epsilon'\\
Q_2 &= \epsilon'_y\epsilon_y - \epsilon'_x\epsilon_x,\\ 
Q_3 &= \frac{1}{\sqrt{3}}(\vec{\epsilon'}\cdot\vec{\epsilon} -
3\epsilon'_z\epsilon_z),\\
T_x &= \epsilon'_y\epsilon_z + \epsilon'_z\epsilon_y,\\ 
T_y &= \epsilon'_x\epsilon_z + \epsilon'_z\epsilon_x,\\ 
T_z &= \epsilon'_y\epsilon_x + \epsilon'_x\epsilon_y.
\end{align}
We approximate the normal incidence scattering geometry in Fig.~2b by taking $\phi$=$90^\circ$.
%in which case the incoming polarization simplifies to $\epsilon=(1,0,0)$ and outgoing polarizations are
%$\epsilon'=(0,0,-1)$ and $\epsilon'=(0,1,0)$. 
The resulting spectra are
shown in Supplementary Fig.~1a. A comparison to the orbital-resolved spectra shown in
Supplementary Figs.~1c and 1d reveals that the spectral weight is dominated by the {\rm B} doublet. For the $\phi$=$0^\circ$ geometry 
%with $\epsilon=(0,0,1)$ and $\epsilon'=(0,1,0)$ resp. $\epsilon'=(1,0,0)$,
the $\beta$-factors cancel out and the resulting spectrum,
shown in Supplementary Fig.~1b, shows only the higher-energy C mode.

\end{document}